\documentclass[10pt,conference]{IEEEtran}
\usepackage{cite}
\usepackage{amsmath,amssymb,amsfonts}
\usepackage{algorithmic}
\usepackage{graphicx}
\usepackage{textcomp}
\usepackage{xcolor}
\usepackage[hyphens]{url}
\usepackage{fancyhdr}
\usepackage{hyperref}

\newcommand{\hpcayear}{2025}

\graphicspath{{images/}}

\usepackage{subfiles}

\usepackage{xspace}
\def \projectName {Examem\xspace}

\usepackage{listings}

\AtBeginDocument{%
  \providecommand\BibTeX{{%
    \normalfont B\kern-0.5em{\scshape i\kern-0.25em b}\kern-0.8em\TeX}}}

\newcommand{\mnote}[1]{}
\newcommand{\anote}[1]{}
\newcommand{\hnote}[1]{}

\newcommand{\hpcasubmissionnumber}{321}
\title{\projectName: Low-Overhead Memory Instrumentation for Intelligent Memory Systems}

\def\hpcacameraready{} 

\newcommand\hpcaauthors{Hayden Coffey$^1$ Ashwin Poduval$^1$ Michael Swift}
\newcommand\hpcaemail{}
\newcommand\hpcaaffiliation{University of Wisconsin-Madison}

\author{
  \ifdefined\hpcacameraready
    \IEEEauthorblockN{\hpcaauthors{}}
      \IEEEauthorblockA{
        \hpcaaffiliation{} \\
        \hpcaemail{}
      }
  \else
    \IEEEauthorblockN{\normalsize{HPCA \hpcayear{} Submission
      \textbf{\#\hpcasubmissionnumber{}}} \\
      \IEEEauthorblockA{
        Confidential Draft \\
        Do NOT Distribute!!
      }
    }
  \fi 
}

\fancypagestyle{camerareadyfirstpage}{%
  \fancyhead{}
  
  \fancyhead[C]{
    \ifdefined\aeopen
    \else
      \ifdefined\aereviewed
      \else
      \ifdefined\aereproduced
      \else
    \fi 
    \fi 
    \fi 
    \ifdefined\aeopen 
      \includegraphics[width=12mm,height=12mm]{ae-badges/open-research-objects.pdf}
    \fi 
    \ifdefined\aereviewed
      \includegraphics[width=12mm,height=12mm]{ae-badges/research-objects-reviewed.pdf}
    \fi 
    \ifdefined\aereproduced
      \includegraphics[width=12mm,height=12mm]{ae-badges/results-reproduced.pdf}
    \fi
  }
  \fancyfoot[C]{}
}
\begin{document}
\maketitle

\ifdefined\hpcacameraready 
  \thispagestyle{camerareadyfirstpage}
  \pagestyle{empty}
\else
  \thispagestyle{plain}
  \pagestyle{plain}
\fi

\newcommand{\hpcaheight}{0mm}
\ifdefined\eaopen
\renewcommand{\hpcaheight}{12mm}
\fi



\begin{abstract}
   Memory performance is often the main bottleneck in modern computing systems. In recent years, researchers have attempted to scale the memory wall by leveraging new technology such as CXL, HBM,  and in- and near-memory processing. Developers optimizing for such hardware need to understand how target applications perform to fully take advantage of these systems. Existing software and hardware performance introspection techniques are ill-suited for this purpose due to one or more of the following factors: coarse-grained measurement, inability to offer data needed to debug key issues, high runtime overhead, and hardware dependence. The heightened integration between compute and memory in many proposed systems offers an opportunity to extend compiler support for this purpose. 
   
   We have developed \projectName, a memory performance introspection framework based on the LLVM compiler infrastructure. \projectName supports developer annotated regions in code, allowing for targeted instrumentation of kernels. \projectName supports hardware performance counters when available, in addition to software instrumentation. It statically records information about the instruction mix of the code and adds dynamic instrumentation to produce estimated memory bandwidth for an instrumented region at runtime. This combined approach keeps runtime overhead low while remaining accurate, with a geomean overhead under 10\% and a geomean byte accuracy of 93\%. Finally, our instrumentation is performed using an LLVM IR pass, which is target agnostic, and we have applied it to four ISAs.
\end{abstract}

\maketitle
\pagestyle{plain}

\footnotetext[1]{Both authors contributed equally to this work}

\section{Introduction}
\documentclass[../main.tex]{subfiles}
\graphicspath{{\subfix{../images/}}}

\begin{document}

Data movement is a major system bottleneck~\cite{Google-Workloads-Consumer}, dominating runtime overheads and energy consumption.  A variety of approaches have been proposed and used to soften the memory wall. Some recent work \cite{Demystifying-CXL} has proposed leveraging CXL to expand memory bandwidth in addition to capacity. Additionally, there exists a rich body of research proposing processing-in-memory (PIM) and near-memory processing (NMP) systems to move computation closer to memory. These systems often make use of 3D stacked DRAM technology like High Bandwidth Memory (HBM) and Hybrid Memory Cube (HMC) with accelerators~\cite{CoNDA,ABNDP,SDAM,TETRIS}.

Such architectures are unlikely to offer performance benefits for entire applications -- rather, developers will need to identify the most data-intensive sections of code in their applications~\cite{CoNDA, PIM-workload-driven-perspective} that achieve significant performance or efficiency gains after accounting for offload overhead incurred. Even CXL-based systems, which do not involve offloading code, require someone (usually the OS) to make decisions about data placement across memory tiers. 

Data-intensive code must take advantage of higher memory bandwidth by generating a large number of parallel memory accesses regardless of how bandwidth is made available. These highly parallel applications frequently make use of vectorization and multithreading~\cite{To-PIM-or-not}, simultaneously issuing many accesses to different memory channels. However, writing code which effectively uses such techniques is challenging. For instance, a developer may write or reuse code that leads to unequal distribution of work across threads, or use data\ structures with synchronization bottlenecks.

\begin{figure}
    \centering
    \includegraphics[width=\linewidth]{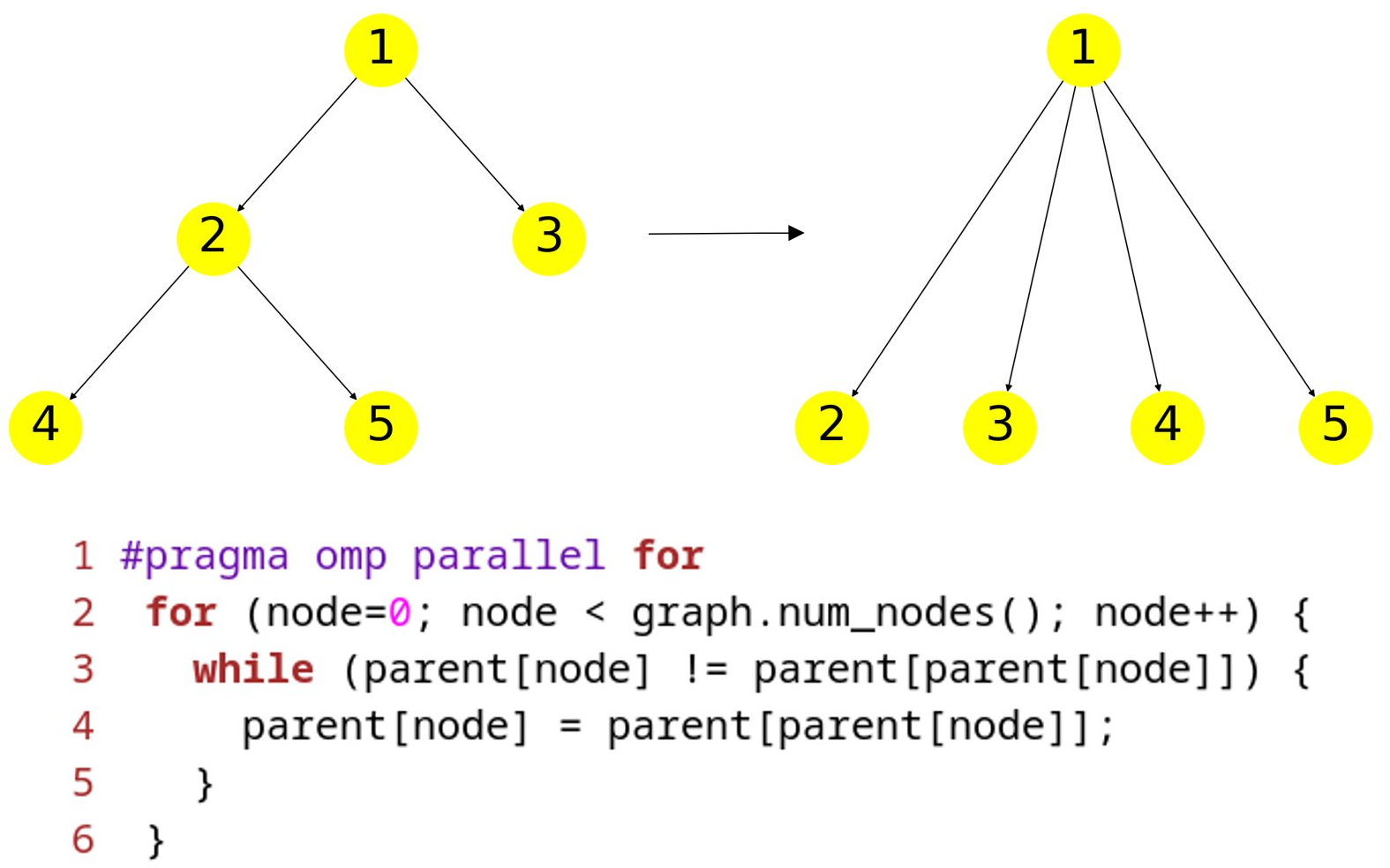}
    \caption{Data-intensive kernel with unequal bandwidth usage. The figure on the left is transformed into the figure on the right by the algorithm. 
    }
    \label{fig:x86-bandwidth}
\end{figure}

Figure~\ref{fig:x86-bandwidth} presents an example kernel based on the shortcutting operation in the Shiloach-Vishkin parallel algorithm~\cite{SHILOACH198257} and depicts its effect. The code snippet involves pointer chasing for each node to transform each tree to which a node belongs into one of unit height (lines 3 and 4.) 
With five OpenMP threads for the example graph, OpenMP's static scheduling policy assigns thread 1 to iteration 1, thread 2 to iteration 2 etc, with each thread assigned to a core with private L1 cache. Per line 3, threads 1, 2 and 3 will make 1 memory access (since their parent and grandparent nodes are the same), while, threads 4 and 5 will need to make 2 accesses, leading to unequal bandwidth usage across threads.

Workload imbalance across cores is but one factor that may prevent an application from optimally using bandwidth. Some other factors that could lead to the same include inefficient data structure layouts, and poor memory placement in NUMA systems. Programmers leverage processor performance montioring units (PMUs) in high performance CPUs to trace events in hardware to debug these kind of issues.

However, these counters are not available in all environments. Virtualized environments may be configured to limit access to hardware performance counters for security \cite{intel-tdx-arch-spec}\mnote{Add citation}\anote{I couldn't find a paper discussing when it is limited, but the capability exists and is mentioned in Intel's documentation, citing that}. Existing performance introspection frameworks like Pin~\cite{Pin} or perf~\cite{perf} are built for specific target architectures or assume OS and hardware performance monitoring unit (PMU) support, with little to no support for new device architectures and ISAs. For example, perf support on RISC-V systems is still a work in progress ~\cite{Perf-improvements-RISC-V}. Moreover, PMUs in ASICs may be limited or missing due to resource limitations.

Identifying such issues require tools that can analyze the most data-intensive subsections of applications.
High-level tools only provide function timing and call stack information, as with gprof~\cite{Gprof} and XRay~\cite{XRay}. Tools like PISA~\cite{PISA} offer more information without hardware support through heavy binary instrumentation, but slow execution by several orders of magnitude. As a result, they can measure what code does, but not how hardware performs. Pin and gem5~\cite{gem5} can monitor specific regions in code, but have high overhead. 


We introduce \projectName, a low-overhead framework for memory performance introspection at the granularity of developer annotated subsets of programs, which we term {\em regions of interest} (ROI). We have three high-level goals for \projectName: first, we seek a system capable of instrumenting arbitrary parts of a program with low runtime overhead to be suitable for continuous profiling or instrumenting offloadable code. Second, we seek  portability to a wide range of ISAs and accelerators by minimizing dependence on archictecture-specific features. Finally, we want to instrument multi-threaded workloads and generate per-thread measurements.
\projectName recognizes regions of interest and collects instruction information at compile time to reduce overhead. If performance counters are available, it can snapshot them before and after a ROI and store the difference. \projectName implements a basic-block profiler that adds instrumentation to record time spent in a region of interest. It adds counters that are incremented at runtime and are used in combination with the information generated during compilation to measure the memory bandwidth and instructions executed for a region of interest. Furthermore, we implement optimizations to reduce the amount of instrumentation to less than one per basic block. 

We implement \projectName on LLVM, allowing it to be ported to a wide variety of devices, with the majority of its code being implemented in an IR pass, which is architecture agnostic. It supports multi-threaded execution with per-thread counters, and uses sampling to reduce overhead. While we focus on memory bandwidth, \projectName can also record other information about ROI execution.

We evaluate \projectName on 24 applications from the GAPBS~\cite{beamer2017gap} and SPEC2017~\cite{SPEC2017} benchmark suites.  \projectName instrumentation increases geomean execution time by 8\% compared to uninstrumented binaries with a sampling rate of 1\%. In comparison, pintool runtime is 5-7x longer. 
We compare counts and data volumes of loads and stores for these benchmark suites against a pintool, and find \projectName's accuracy is about 93\% for data volume (bytes accessed) and 87\% for bandwidth. We demonstrate \projectName's portability with an evaluation on an AArch64 system, and additionally confirm its functionality on RISC-V and POWER8 systems. \mnote{Add RiscV}\anote{added it and power isa - we only confirmed functionality for all three, but we could only compare arm64 against x86}
%

We make the following contributions in this paper:
\begin{enumerate}
    \item We introduce \projectName, a novel memory performance introspection framework providing detailed performance information about user-annotated regions of interest.
    \item We describe an optimized instrumentation process that can accurately obtain information about the instruction mix and memory bandwidth for regions of interest with low overhead.
    \item \anote{Shortening by not mentioning all ISAs by name}We show that \projectName is easy to use on new architectures by deploying it on four ISAs.
\end{enumerate}





\section{Overview}
\documentclass[../main.tex]{subfiles}
\graphicspath{{\subfix{../images/}}}

\begin{document}

We designed \projectName as a target-independent memory performance introspection tool to support a variety of use cases and hardware environments. Specifically, we designed \projectName to accurately report information about memory operations and their performance both at the granularity of a region being instrumented as well as for its component basic blocks for more fine grained analysis and tuning of program behavior. 

With the rise of domain specific computing and the plethora of architectures being proposed to tackle the memory wall, we seek an approach that minimizes the effort to port it to a new architecture. 
With parallel processing needed to fully saturate memory bandwidth, instrumentation must be thread safe. Finally, instrumentation must be lightweight to minimize the impact on achieved memory bandwidth if accurate bandwidth estimates are to be made.


\begin{figure}
    \centering
    \includegraphics[width=\linewidth]{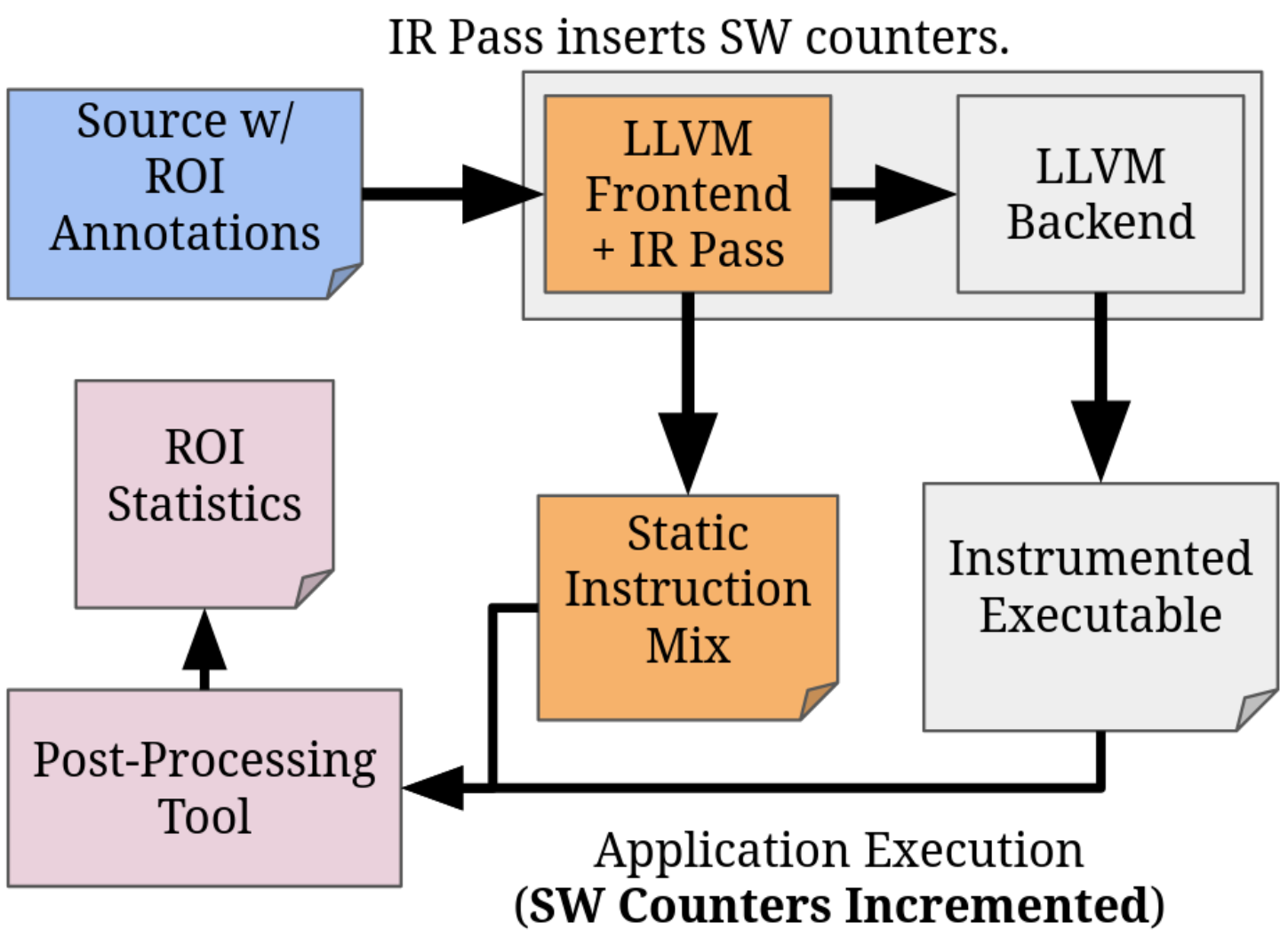}
    \caption{Overview of \projectName's components.}
    \label{fig:zray_general_overview}
\end{figure}

\projectName is comprised of an LLVM IR (intermediate representation) pass and a post-processing tool. Figure \ref{fig:zray_general_overview}
illustrates the general relationship between these components as an application is profiled. To instrument an application with \projectName, users mark the regions of interest (ROI)
they wish to profile using pragma statements to specify the begin and end boundaries. A developer may use one of many techniques described in literature to identify data-intensive code. This could either involve profiling applications and identifying how frequently they access memory, using stats like L1 and LLC miss rates~\cite{DAMOV}, or device-specific heuristics~\cite{To-PIM-or-not} which take into account the cost of offload and the potential speedup from offloading the ROI code to an accelerator.

In addition to profiling ROIs, \projectName supports instrumenting all basic blocks in the program, unlike standard profiling tools like gprof that instrument functions only. This can be used to help identify the most frequently accessed basic blocks and functions, then mark ROIs accordingly as well as those that are most memory intensive and likely to benefit from offload.

After annotating ROI, users compile their application to an IR file format that \projectName can operate on. \projectName's LLVM pass inserts instrumentation into the IR file, from which executable code is generated. 

\begin{figure}
    \centering
    \includegraphics[width=0.75\linewidth]{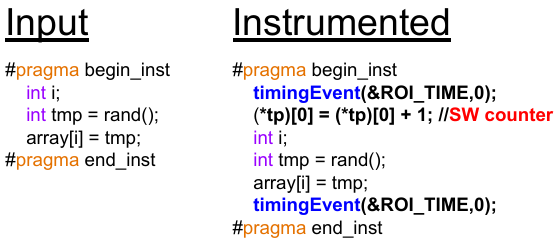}
    \caption{\projectName applies timing events and software counters to regions of interest.}
    \label{fig:zray_c_to_c}
\end{figure}

\projectName's IR pass identifies ROI and performs control flow analysis to determine optimal locations in the ROI to insert software counters for the block profiler. The pass inserts code to statically allocate software counters at program start, with counter increment and timing instructions inserted into the ROI at the identified insertion points as shown in Figure~\ref{fig:zray_c_to_c}. In environments where hardware performance counters are available, \projectName enables users to snapshot hardware performance counters before and after an ROI along with timing data. 

The post-processing routine runs after application termination and produces output files containing
the dynamic instruction mix, timing information, achieved memory bandwidth, and performance counter stats (when available) per thread for every ROI. Additionally, the post-processing routine also logs counter, instruction and memory statistics per thread for every basic block instrumented to a CSV file for user analysis with software like pandas or R. \projectName can optionally periodically dump stats to the CSV file to provide time series information, instead of aggregated stats at the end of execution. The post-processing routine calculates these stats by combining software counter values with the instruction mix information recorded at compile time. The current implementation of \projectName does not model the cache hierarchy for memory accesses and reports {\emph{achieved}} memory bandwidth not bandwidth to DRAM.
\anote{Cut a line here. I think it's hard to comment about locality unless the programmer has knowledge before hand about application behavior.}

Overall, we see three use cases for \projectName:
\begin{enumerate}
\item For applications bottlenecked by memory, identify and profile regions of code responsible for such bottlenecks and make algorithmic optimizations
\item While developing code for NMP or PIM devices to identify offload targets as discussed previously
\item On NMP or PIM devices, once they are made available, to measure achieved performance. 
\end{enumerate}






\section{Design}
\documentclass[../main.tex]{subfiles}
\usepackage{graphicx}
\graphicspath{{\subfix{../images/}}}

\begin{document}

We  now describe each component of \projectName in more detail, as well as a set of optimizations
implemented to improve profile accuracy and reduce instrumentation overhead.


\subsection{IR Pass}
The IR pass is responsible for identifying code within regions of interest and statically analyzing the basic blocks that comprise them. LLVM provides a rich suite of analysis passes that can be used to organize the control flow graph (CFG) and identify useful properties such as domination and loop iteration counts. 

During compilation, clang translates inline assembly comments marking the beginning and end of a ROI. \projectName identifies basic blocks belonging to the ROI by using these comments to filter out a subset of the CFG for further analysis. Additionally, since a ROI may only contain a subset of a basic block, the IR pass splits the basic blocks containing the ROI begin and end statements to prevent the instrumentation of unrelated instructions.

Once the blocks belonging to the ROI are selected, \projectName constructs a basic block profiler
with lightweight and thread local software counters. Each counter is associated with a set of basic blocks
and records the number of times the first block of that set is executed. 
In order to guarantee that each block in the set is executed together,
\projectName constructs the sets using a post-domination analysis. That is, by examining the CFG structure, \projectName groups basic blocks
together such that if the first block in a set is executed, the subsequent blocks in that set are guaranteed to also be executed. We elaborate more on how these sets are constructed in Section
\ref{sec:design-opt}.

These post-domination sets are a variant of
super blocks introduced by Agrawal~\cite{agrawal1994dominators}.
However, Agrawal was targeting
code coverage, while \projectName needs execution counts, requiring a mechanism
to efficiently track the number of times each block is run. 
This is complicated by the potential presence of cycles in the CFG which can result in varying execution counts
for blocks in the same set. We explain how \projectName resolves this challenge in Section \ref{sec:instrument_loop}.

\subsection{Counter management}
\projectName's IR pass inserts additional bookkeeping instructions
responsible for managing the software counters, ensuring the
correct counters are incremented as the application executes. 
The IR pass inserts function calls at the beginning and end of each ROI. These functions serve three purposes. First, they add timing instructions to keep a running total of time spent in each ROI. Second, they include control logic to enable or disable counters for the region. This keeps track of the number of times an ROI is hit and enable counters every Nth time for a sampling period of N.

Finally, the instrumentation reads hardware counters in addition to the time stamps when hardware counters are available and enabled in \projectName. We use the perf event library and read L1 misses, LLC misses, and total instructions executed by default. However, users can update the handler, which consists of less than 50 lines of C++ code, to leverage other methods of accessing counters or to record other events. We add counter reads at the start and end of each ROI to measure the total stats for the ROI. These reads are placed outside the timing events, before the starting timer call and after the ending timer call to ensure that the overhead associated with a perf event read does not pollute ROI timing measurements.

The IR pass statically allocates the software counters as an array in thread-local memory.
It also inserts inlinable function calls to update counters for each set of basic blocks. These functions accept arguments that are used to index into the counter array and increment the corresponding value.
Additional care must be taken when multiple ROI are involved with shared functions to ensure that
execution of a function from within one ROI does not impact the profiles of the other ROI. We describe
this challenge and how \projectName resolves it in more detail in Section \ref{sec:functions}. 

Lastly, on application or thread termination, inserted code writes the software counter values and ROI statistics out to a file for later post processing.  

\subsection{Recorded static event types}
\mnote{Rename section once written. Write something about what events we can count - what is default, what can be done, what is hard (e.g., arguments as can't aggregate over loops)}\anote{Done}
\projectName counts instruction-level events within ROI. By default, it counts the number of instructions that access memory and the size of the argument, allowing it to compute the total bytes accessed. As instrumentation is done in an LLVM IR pass, which is architecture agnostic, local variable and parameter access is represented by variables that may be on the stack or in registers. \projectName tracks all memory instructions it can find, but can only guarantee accurate counts for those referencing heap memory and global variables\mnote{What about globals?}\anote{Mentioned, updated this section}. This is because many stack accesses are due to register spilling or saving across function calls, which are not visible at the IR level.
\mnote{Fix up and add more - e.g. counting FP operations}

In addition to memory accesses, \projectName counts instruction-level events visible to the IR, such as integer and floating point operations, and control flow instructions like jumps, branches, and calls. However, it cannot count events related to instruction parameters, such as addresses or values, as they aren't available statically and can change at runtime, which would require instrumenting and recording information about every execution of an instruction.

\subsection{Minimizing Counters with Post-Domination Sets}
\label{sec:design-opt}
\begin{figure}
    \centering
    \includegraphics[width=1\linewidth]{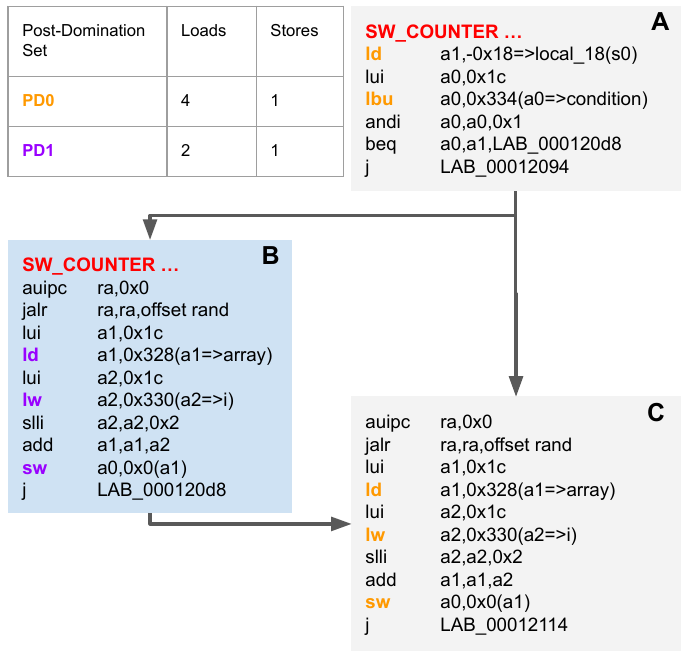}
    \caption{Example post-domination set. Block C post-dominates
    block A; one counter covers the two blocks. Block B does
    not post-dominate A, and is given its own counter.}
    \label{fig:postdomset}
\end{figure}

If the goal is to capture all memory instructions  performed in the ROI of the application, a simple
approach would be to maintain counters for these instruction types and increment them on every memory
instruction execution. This can easily be done using Intel's Pintool~\cite{Pin} by examining
each executed instruction at the cost of significant
instrumentation overhead and an architectural dependency on x86-64.

By recognizing the simple properties of basic blocks (single entry/exit points),
a single software counter in combination with a collected data profile for the basic block
can account for all instructions within the block as they will be executed together
barring any exceptions or failures. \projectName further {\em merges} the instrumentation of different basic blocks
in the CFG together as shown in Figure \ref{fig:postdomset}. \projectName tests for the property of post-domination to create sets of blocks as a variant of super
blocks~\cite{agrawal1994dominators} where each block is guaranteed to be
executed together. By grouping basic blocks into post-domination sets, one software counter
can now cover multiple basic blocks.

\projectName uses post-domination sets to minimize the number
of software counters needed to cover the basic blocks of the ROI. 
When creating the post-domination sets, \projectName's IR pass will select the first
basic block in the ROI and add it to a set with other basic blocks that post-dominate it.
It repeats this process with the remaining basic blocks that did not post-dominate
the first selected basic block until all basic blocks are assigned a set. With the
sets constructed, the IR pass allocates a software counter for each post-domination set,
and inserts counter increment instructions into the first basic block of each set.
Then, the IR pass records the static instruction mix information of the basic blocks comprising each set and writes this information to the static mix file for use by the post-processing tool.

\subsection{Instrumenting Loops}
\label{sec:instrument_loop}

\begin{figure}
    \centering
    \includegraphics[width=1\linewidth]{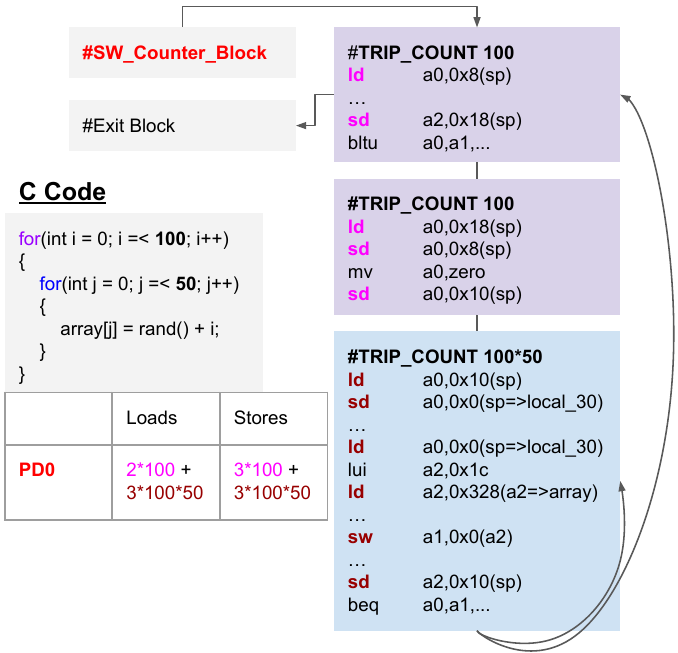}
    \caption{Example loop SW counter hoisting. One counter covers both for loops
    by scaling the instruction mixes of the basic blocks by the
    loop trip counts.}
    \label{fig:loop_hoist}
\end{figure}

\projectName applies post-domination set tracking to ROI involving loops by first separating out basic blocks involved in loops from those executed at most once per CFG execution.
For each loop,
\projectName uses LLVM's analysis framework to identify the loop's trip count
(number of loop iterations to be performed).
If the trip count can be identified at compile time, the static instruction mix
of the post-domination set owned by the loop's header block
are multiplied by the trip count when the IR pass writes them to the static mix file.
The static mix is now representative of the instructions executed
per entire loop execution.
For improved performance on loops with identified trip counts,
\projectName hoists the respective software
counter above the header block to a new basic block so that it is only executed once per entire loop execution. 
Blocks that do not post-dominate the header (e.g., conditional statements within the loop)
are treated as not belonging to a cycle and are instrumented normally.

If the trip count can only be determined at runtime (e.g., controlled by a variable), \projectName's instrumentation uses the dynamic trip count to update the loop header's count in one operation, rather than incrementing the count once each executed iteration.
If LLVM is unable to calculate the loop trip count
at either compile or runtime (e.g., the trip count is not loop invariant),
\projectName falls back to instrumenting all of the loop's blocks without any scaling
applied, requiring a counter increment for each loop iteration. 

A complication in this scheme arises when nested loops are involved as shown
in Figure \ref{fig:loop_hoist}, where we have a parent and child for loop.
While one option would be to simply parse each loop individually, if
both loops' trip counts can be calculated at compile time, we can combine the trip
counts from each loop by applying 
the parent's trip count to its unique blocks and the product of the parent and child's trip counts to the child's basic blocks. With
this adjustment, a single software counter can cover basic blocks across both loops and reduce instrumentation overhead substantially.
We can continue this process and hoist software counters further up the CFG so long as each loop's trip count can be calculated at compile time and
the parent loop's header is post-dominated by the subsequent child loop's header. 

\subsection{Functions and External Code}
\label{sec:functions}
ROI may spend most of their runtime in called functions, executing instructions not explicitly
within the ROI code. 
Ignoring function calls sacrifices too much accuracy in most programs,
and instrumenting all called functions will cause overhead on code paths 
outside the ROI as well as pollute the instrumentation
logs. 

To address these issues, \projectName's IR pass uses {\em function cloning} to create copies of
functions called within ROI and applies instrumentation to these clones. For example, if
ROI $A$ calls function $B$, \projectName will create a function clone $B^*$ with instrumentation and replace
all function calls to $B$ inside $A$ with $B^*$. This cloning is only done once and additional ROI
calling $B$ will be directed to $B^*$ as well. 
However, the challenge still remains in distinguishing counter increments when
multiple ROI call the same cloned function. As an example, suppose ROI $A$ and ROI $C$ both
use $B^*$, an instrumented clone that requires 5 counters to record accesses to its basic blocks. How can \projectName ensure that $A$ calling $B^*$ will only update its own information
and not impact the collected
profile of $C$?
\projectName resolves
this challenge by allocating a separate set of counters corresponding to a cloned function for each ROI during the
IR pass -- in this example it allocate 5 counters each to $A$ and $C$ for recording accesses in $B^*$. \projectName inserts bookkeeping instructions to
select the appropriate set of counters to update when first entering
a ROI. 

Function cloning only works if \projectName has access to the function definition during the IR pass. At the moment, there are two cases that \projectName cannot automatically handle: indirect functions and external library functions. For the first case, the developer can manually determine which functions are likely to be called and instrument them, analogous to manually performing function cloning. For example, if an indirect function call in an instrumented function $A$ can only call functions $B$ or $C$, the developer can also add ROIs to $B$ and $C$ to ensure that all accesses of interest will be recorded. For the second, a developer could compile the library code down to IR and merge it with the application's IR with llvm-link before applying \projectName's pass to it. Any functions called from the linked library will now be instrumented via function cloning.

LLVM uses intrinsics for some standard library functions, particularly the memory movement/setting functions  \texttt{memcpy}, \texttt{memmove}, and \texttt{memset} are critical for accurate measurements.
\projectName has a dedicated mechanism to account for these intrinsics. The IR pass records the size of the intrinsic operation, if known at compile time, to the static mix file.
However, the operation size may be unknown at compile time. In such cases, the IR pass can identify which variable holds
the operation size. Since we can only determine the size of these operations at runtime, \projectName adds instructions at compile time to use this variable at runtime to record the number of bytes accessed, similar to counter array instrumentation for basic block counts.
To avoid aggregating intrinsic access sizes with basic block counts, \projectName allocates separate data structures to record these unknown intrinsic sizes, the load and store intrinsic size array.


\projectName allocates and accesses the intrinsic size array in a manner similar to the counter array and all intrinsics in a basic block or set of basic blocks update the same intrinsic size array value. Memmove and memcpy involve reads and writes increment operations to both load and store arrays; memset only updates the store intrinsic array. 

This approach allows us to cheaply and accurately record the number of  {\em bytes read from or written to} memory for these instructions. However, under the hood, the standard library may perform transfers at any granularity of accesses, with potentially varying sizes, e.g., 4 or 8 bytes. \projectName has no way of determining this and assumes 8-byte operations. This approach underestimates the count of {\em memory accesses} when transfers are performed at smaller granularity.

\subsection{Post-Processing Tool}
\label{sec:post-process}
The post-processing tool combines static information 
(instruction mix, bytes accessed)
collected by the IR pass with dynamic information
(software counters, timing information)
collected by the inserted instrumentation at runtime.
The final logged results comprise: counter values, dynamic instruction mix, region timing information,
and achieved memory bandwidth estimates. 

The post-processing tool calculates
dynamic instruction
mix by multiplying the static instruction mix for each post-domination set with its
respective software counter value. The post-processing tool also adds the number of bytes accessed and estimated count of memory access operations for intrinsics from the intrinsic size array to the dynamic mix at this point. The number of bytes read and written by the ROI is divided by ROI timing information to estimate achieved read and write memory bandwidths respectively. 

However, following this naive approach leads to \projectName reporting lower memory bandwidth than an ideal profiler would due to counter instrumentation
overhead, which slows ROI execution.
We account for instrumentation overhead using a heuristic {\em correction factor} ($CF$) derived from the increase in instruction count due to counters.

Each counter increment involves a read-modify-write. Memory operations are generally among the most expensive, with even L1 accesses being several times more expensive than ALU operations. To account for this, we scale the counter instruction count in our correction factor by a memory factor, derived from the original memory access fraction (OMAF) of the ROI before counters were inserted and the new memory access fraction (NMAF) after inserting counters. Optimizing this heuristic is an option for further improvement on \projectName.
\begin{equation}
\text{OMAF}=\frac{\text{Total Memory Accesses}}{\text{Instructions Exec.}}
\vspace{2 mm}
\end{equation}
\begin{equation}
\text{NMAF}=\frac{\text{No. Counter Incr.}*2 + \text{Total Memory Accesses}}{\text{No. Counter Incr.}*2 + 
\text{Instructions Exec.}} \\
\end{equation}
\begin{equation}
\text{Memory Factor (MF)}=\frac{\text{NMAF}}{\text{OMAF}} \\
\end{equation}
\begin{equation}
CF=\frac{\text{Instructions Exec.}  + (\text{No. Counter Incr.} * \text{MF})}{\text{Instructions Exec.}}
\end{equation}


Finally memory access bandwidths are reported as follows:
\begin{equation}
    \text{ROI Read/Write BW} = CF * \frac{\text{ROI Bytes Read/Written}}{\text{ROI Total Time}}
\end{equation}



\begin{figure}
    \centering
    \vspace{1.75 mm}
    \includegraphics[width=1\linewidth]{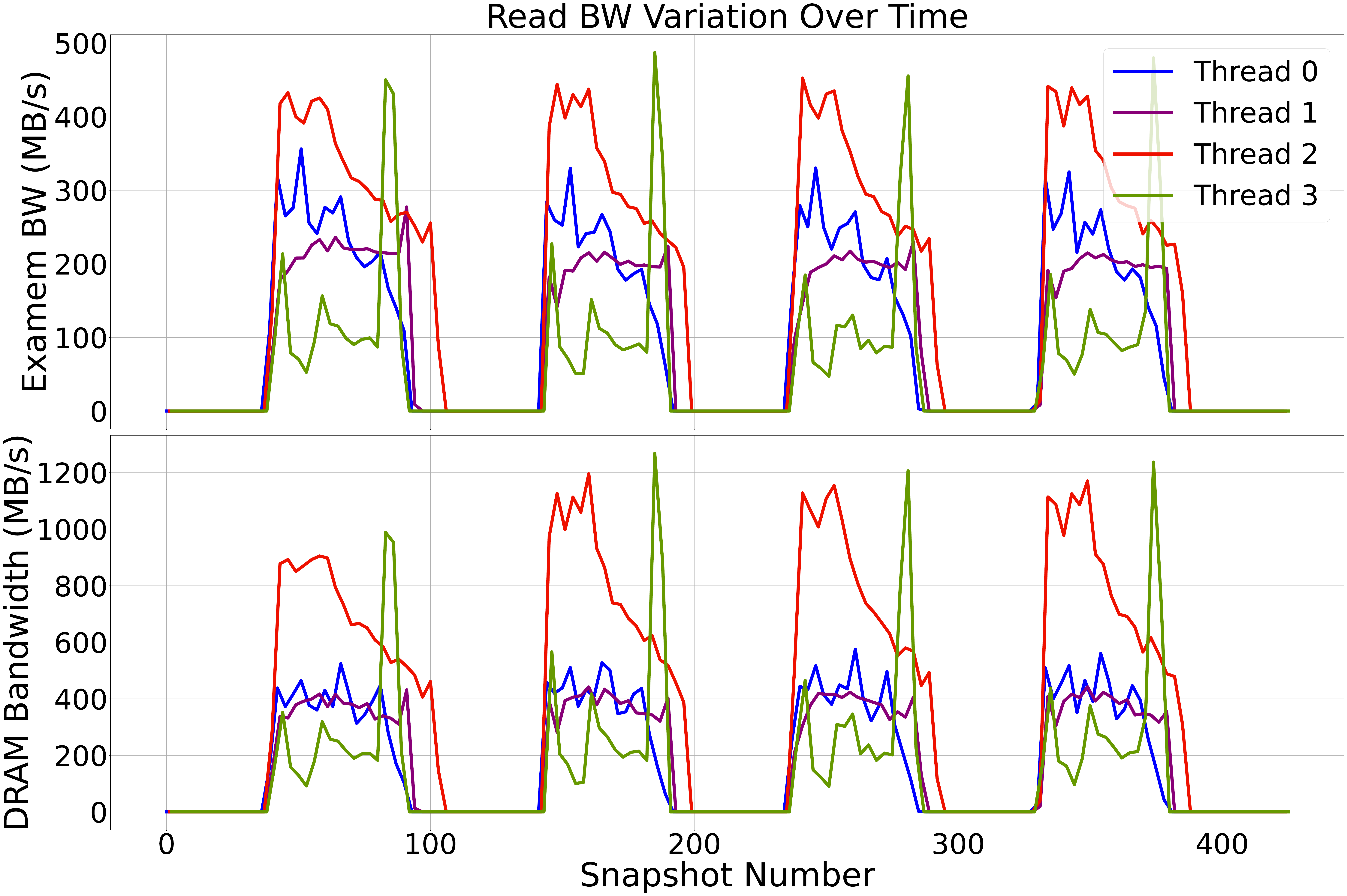}
    \caption{Time series plot of read bandwidth for the Shiloach-Vishkin connectivity algorithm as reported by \projectName (upper plot) and DRAM bandwidth (lower plot) calculated using the number of LLC misses. X-axis is the snapshot number, analogous to time. It is the polling period to which the corresponding bandwidth measurement belongs to. }
    \label{fig:bw-time-series}
\end{figure}

\subsection{Runtime Performance Monitoring}
Waiting until program termination to check memory performance is impractical for long running or continuous applications. To address these scenarios, we built a message-based monitor on top of \projectName. \projectName-instrumented threads perform a handshake with the monitoring process on startup to share the thread's ROI memory addresses. The monitoring process periodically polls counter values and executes the post-processing logic to calculate performance statistics for each epoch. Plotting these statistics over time can reveal phases in program execution, as well as thread imbalance, with an example shown in Figure \ref{fig:bw-time-series}. 

In this example, we study the bandwidth for an implementation of the Shiloach-Vishkin connected components calculation algorithm in the GAPBS benchmark suite, a data-intensive workload with a high LLC MPKI. We configure OpenMP to spawn 4 threads for simplifying the visualization. We record read bandwidth as captured by \projectName. Additionally, we enable \projectName's optional hardware performance monitoring and record LLC miss information to calculate DRAM bandwidth usage.

\projectName's bandwidth measurements closely match the behavior and trends followed by actual DRAM bandwidth for every thread monitored. The real DRAM bandwidth numbers in this case, surprisingly, are somewhat higher than those reported by \projectName. One reason for this could be that the LLC reads from DRAM at the granularity of 64 bytes. On the other hand, \projectName measures the accuracy of heap accesses to L1 keeping in mind the exact size of data type and data requested, which is usually 4 or 8 bytes. The workload involves extensive pointer chasing over neighbors of each node. Thus, it is likely that each cache line is accessed only one or a few times. In essence, we are fetching 64 bytes from DRAM for \textless 64B of read operations.

We tested \projectName with other workloads in the GAPBS \cite{beamer2017gap} suite and found that it similarly captured information about the bandwidth behavior for other workloads with high LLC MPKIs. For more cache-friendly workloads with lower MPKI, \projectName is no longer able to capture the behavior, but it does identify trends such as sharp rises or drops in bandwidth usage.

Any interface that allows for memory access can support this form of monitoring, such as remote monitoring over RDMA, as the monitor is simply copying memory. In practice, we found monitoring to involve copying on the order of tens of KB every couple seconds for large instrumented programs (1000+ counters). Users may adjust the intensity of instrumentation and the polling rate to reduce communication demands or capture more information.

\subsection{Implementation}
\mnote{Added - need to flesh out}
\projectName comprises the LLVM instrumentation pass, runtime library for managing counters, post-processing tool, and communication library for periodically sampling counters.  We report the number of lines of C++ code needed to implement each module of \projectName in Table \ref{loc-table}.

\begin{table}[]
    \centering
    \caption{Lines of code needed for each module of \projectName}
    \begin{tabular}{l|r}
    Component & Lines \\
    \hline
    IR Pass & 2534 \\
    Runtime & 681 \\
    Post-processing & 386 \\
    Communication & 150 \\
    \hline
    \end{tabular}
    \label{loc-table}
\end{table}

\section{Evaluation}
\documentclass[../main.tex]{subfiles}
\graphicspath{{\subfix{../images/}}}
\begin{document}

We answer the following questions in this evaluation:
\begin{enumerate}
    \item What is \projectName's accuracy for reported bytes read and written by heap accesses? 
    \item What is the accuracy of heap access  bandwidth reported by \projectName?
    \item What is the overhead of \projectName, and how effective is sampling in reducing overhead?
    \item What is the effort to port \projectName to another architecture?
\end{enumerate}


We determine \projectName's heap access accuracy to evaluate its ability to capture information for operations which are likely to result in DRAM accesses, because stack accesses have high cache hit rates. Low overhead ensures developers receive timely results, and allow it to be used for continuous profiling.

We developed \projectName to leverage compute available on new memory devices. For this reason, it needs to be portable and have few (if any) dependencies on the target architecture. We examine \projectName's portability with a study examining the process of porting \projectName to an AArch64 server.

\subsection{Experimental setup}

\paragraph{Compiler} We use a custom version of LLVM-15, with custom pragmas to demarcate the beginning and end of regions of interest.\mnote{Fix preceding sentence - it is unclear what we mean about pragmas}\anote{We did most of our testing with our version of clang, which had custom pragmas to mark the beginning and ending of ROI (and for a long time, we also had the MIR pass) and that is the one we have best accuracy with. With the package manager's version of clang, accuracy is mostly unchanged but is worse for a small number of workloads, probably a new optimization that needs to be accounted for} We tested \projectName with apt's version of llvm-15 and clang-15 for the portability experiments using asm comments instead of pragmas.

\paragraph{Baseline} We measured accuracy using a pintool that records stats for developer annotated regions of code as baseline on the x86-64 system. We ran each workload thrice and recorded the arithmetic mean of runtimes for the overhead experiments. We used Pin 3.31 to run the pintool. 
The pintool uses function instrumentation to identify developer annotated ROI. Within these ROI we track heap memory accesses. We use pin's ability to track the base registers for memory operands to filter out accesses using RSP. 
While DynamoRIO \cite{DynamoRIO-bruening2001design} is compatible with AArch64 \cite{aarch64-dbi}, it lacks some support for some of the functionality needed by our pintool for function-level instrumentation.  

We compare the bandwidth reported for a region of interest using \projectName against the baseline bandwidth which is calculated using the bytes accessed in the region of interest as reported by a pintool and recording time spent in the region of interest without instrumentation.

\paragraph{Hardware and environment} We use an sm110p CloudLab\cite{CloudLab-Duplyakin+:ATC19} server with an Intel Xeon Silver 4314 CPU and 128 GB RAM for our accuracy and performance experiments. The server ran Ubuntu 22.04. We evaluated \projectName's portability using an AArch64 m400 CloudLab server with 64GB RAM running Ubuntu 22.04. We compare its stats against a Xeon Silver 4114 (in addition to the Xeon Silver 4314 server) due to their similar L1 cache capacities per-core.
\mnote{Add RiscV and any other architectures}\anote{Done}We also test \projectName's functionality on a CloudLab ibm8335 POWER8 server and a StarFive VisionFive2 RISCV single-board computer.

We test \projectName on general-purpose systems because real in- and near-memory systems are unavailable and simulations are slow, precluding us from evaluating \projectName with anything but the smallest of input sizes. 

\paragraph{Benchmarks} We evaluate \projectName on x86-64 using all benchmarks in the GAPBS workload and all 16 benchmarks in the SPEC2017 suite written exclusively in C or C++ as per their documentation \cite{SPEC-benchmarks}. The GAPBS suite is representative of data-intensive workloads, while the SPEC suite helps us confirm that \projectName is effective at capturing information for a variety of workloads. We used the twitter datasets \cite{twitter-graph-dataset} linked in the GAPBS repository as input to the GAPBS workloads, and used train workload sizes for SPEC.

We compiled the workloads with \texttt{-O2} optimizations. The GAPBS workloads were built with OpenMP directives enabled and run with 32 threads, with SMT enabled. The SPEC workloads were run with SMT disabled. We only used the GAPBS workloads for evaluating portability on the AArch64 system, and were run with 8 threads on the x86-64 server chosen for comparison.

\paragraph{Identifying ROI} We tried to follow the heuristic described in CoNDA \cite{CoNDA} using function LLC MPKI and execution time for identifying data-intensive kernels where possible. We were able to identify functions meeting the criteria in most of the GAPBS workloads. The methodology did not apply to the SPEC benchmark programs; for these, we determined the top 5 functions by execution time and selected frequently called kernels where possible.

\subsection{Heap byte access accuracy}
\begin{figure}[t]
    \centering
    \vspace{0.5 mm}
    \includegraphics[width=\linewidth]{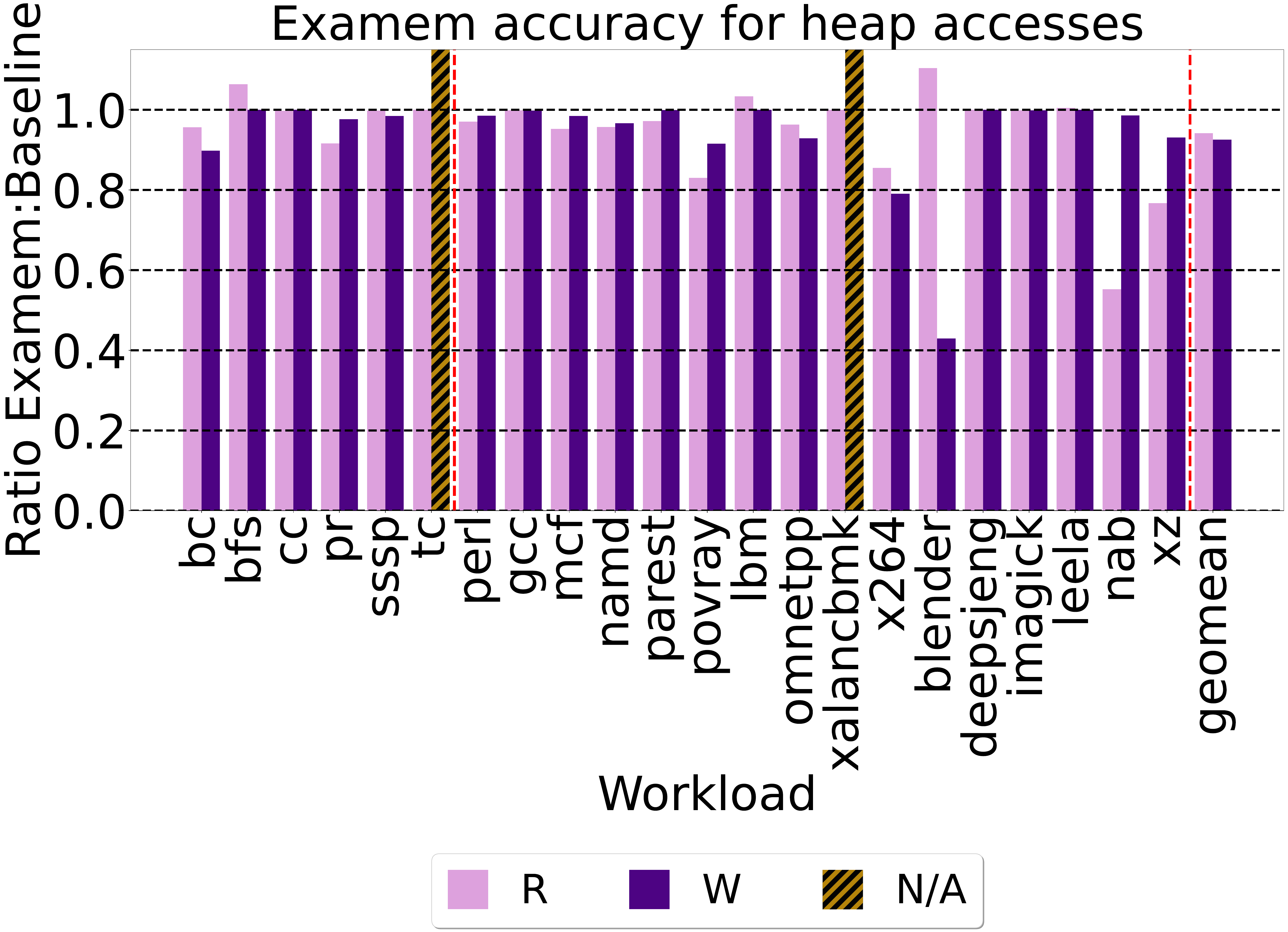}
    \caption{Accuracy of \projectName for total number of bytes read and written during program execution with stack accesses excluded. A value of 1 indicates 100\% accuracy. Values greater and less than 1 represent over- and underestimation of byte counts, respectively. Vertical dotted red lines demarcate different benchmark suites. The last two bars represent geomean across both benchmark suites. 
    }
    \label{fig:ii-byte-wo-stack}
\end{figure}
In this subsection, we examine the accuracy for total number of bytes read from and written to the heap to determine the fraction of user-allocated memory accesses recorded. This is later used to report the read or write bandwidth achieved by an ROI, the accuracy of which is discussed in subsection \ref{bandwidth-acc-subsection}. We do not report the ratio of written bytes for triangle counting and the Xalan C benchmark because both \projectName and the pintool report 0 bytes written for the ROI instrumented. 

As shown in Figure \ref{fig:ii-byte-wo-stack}, \projectName's heap access accuracy is almost perfect for the GAPBS suite, but we do see a few accuracy issues with some of the SPEC workloads. \projectName's accuracy suffers when instrumenting code with many indirect function calls and calls to standard library functions.

Blender is the most notable workload for which we observe the impact of indirect function calls. We instrument the intersect function as the ROI for the workload, which calls RE\_rayobject\_interect. \projectName clones this function. However, a branch within RE\_rayobject\_intersect leads to an indirect call to a RayObject's raycast function, which eventually leads to RE\_rayobject\_instance\_intersect being called.

Our studies indicate that the vast majority of heap accesses within the ROI are reads. Missing a single indirect function that is more likely to write to memory than the rest of the code executed can lead to us missing a large fraction of writes, even if we capture most accesses.

The SPEC nab workload shows poor load accuracy. We instrument the function mme34 in this workload. This inaccuracy results from standard library calls for mathematical operations. The accuracy could be improved by linking called functions from the standard library to IR before running \projectName's IR pass to instrument them via cloning.

Overall, \projectName shows very high accuracy for measuring the count of heap bytes accessed, despite instrumenting the architecture-agnostic IR. This accuracy arises from focusing on heap accesses, which are less architecture specific than stack accesses that reflect register usage differences.

\subsection{Memory bandwidth accuracy} \label{bandwidth-acc-subsection}
\begin{figure}[t]
    \centering
    \vspace{0.5 mm}
    \includegraphics[width=\linewidth]{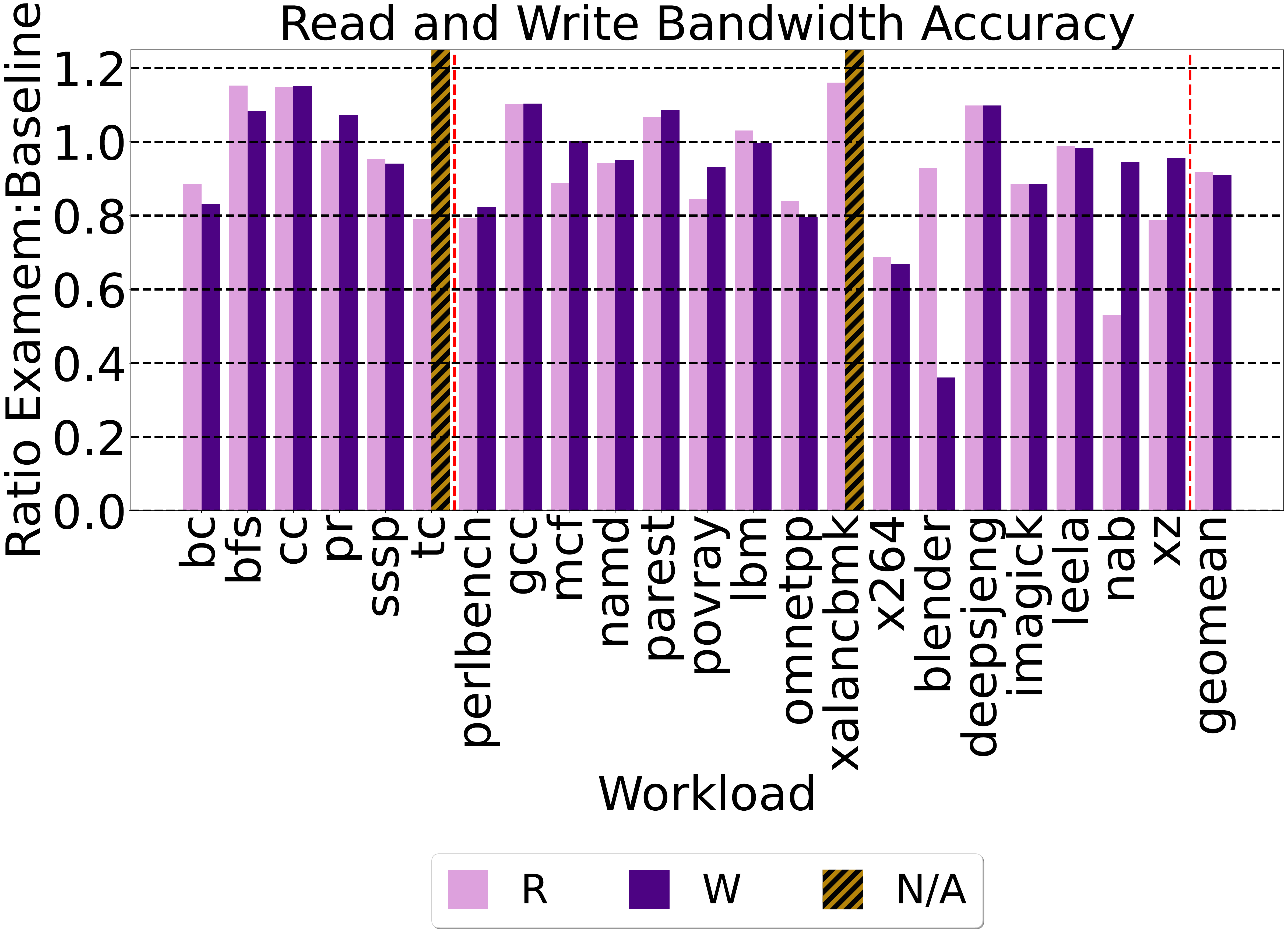}
    \caption{\projectName's read and write bandwidth accuracy. A value of 1 indicates 100\% accuracy. Values greater and less than 1 represent over- and underestimation of bandwidth, respectively. Vertical dotted red lines demarcate different benchmark suites. The last two bars represent geomean across both benchmark suites. 
    }
    \label{fig:i-bandwidth}
\end{figure}
A major goal of \projectName is measuring memory bandwidth. Figure \ref{fig:i-bandwidth} shows the accuracy of heap access bandwidth reported by \projectName against baseline bandwidth. For workloads with multiple ROI, we obtained bandwidth for the largest ROI (in terms of bytes accessed) with both read and write accesses. The GAPBS workloads are multithreaded, and we obtained stats per thread; we obtain bandwidth accuracy by summing the total number of bytes divided by the total time spent for elapsed in each thread ROI.

There is greater variation in reported bandwidth than bytes read/written. The instrumentation adds overhead to the region being monitored.  For most workloads, our correction factor (\ref{sec:post-process}) underestimates the overhead, and \projectName under-reports bandwidth by 10\% on average. Some workloads, such as tc and x264 are affected more strongly. For a handful of workloads, notably about half the GAPBS suite, it leads to reporting bandwidths about 5-15\% higher than baseline. 
Accuracy is also lower for ROI where the total byte count is inaccurate. This issue shows up in blender and nab.

Overall, the bandwidth accuracy is high enough to show important trends, such as in workloads that exhibit phases of execution or bursty behavior, sustained periods of stable bandwidth use, and inefficient or unbalanced memory usage. In addition, the accuracy is enough to show differences across versions of the same program.

\subsection{Overhead}

\begin{figure}[t]
    \centering
    \vspace{0.5 mm}
    \includegraphics[width=\linewidth]{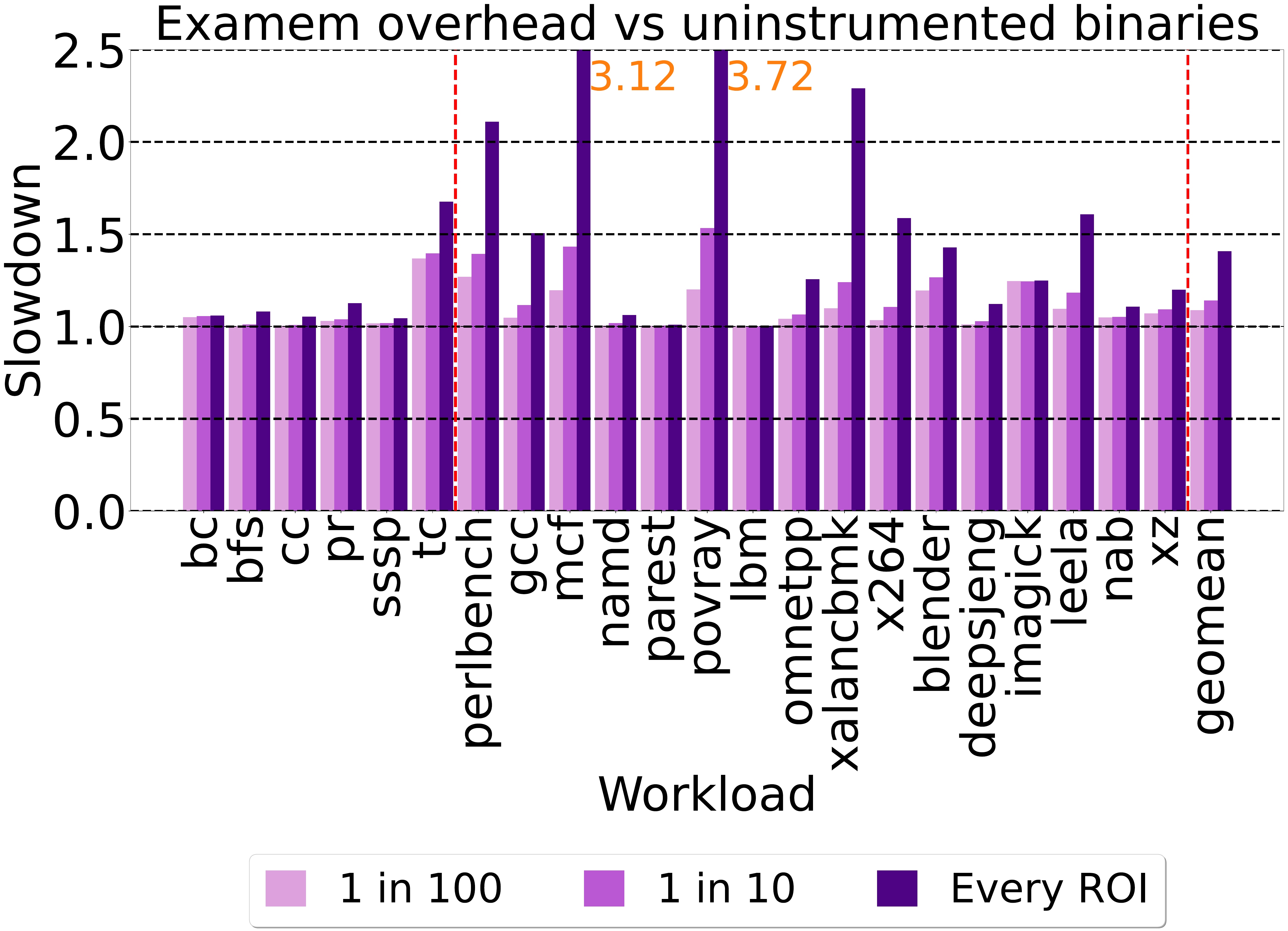}
    \caption{Runtime overhead observed with binaries instrumented using \projectName compared to uninstrumented binaries for different sampling rates. Lower is better - a value of 1 indicates no overhead. Vertical dotted red lines demarcate different benchmark suites. The last three bars represent geomean across both benchmark suites.  
    }
    \label{fig:iii-x86-overhead}
\end{figure}

As shown in Figure \ref{fig:iii-x86-overhead}, \projectName generally has low overhead, with instrumented binaries terminating in a time similar to that of uninstrumented versions. For a few workloads, we observe an overhead of 2x or more when instrumenting every ROI access, as shown in the graph. Running mcf and povray binaries with perf attached indicated that they were core bound after instrumentation, with mcf initially being bottlenecked by bad speculation and povray initially being frontend bound. The workloads becoming core bound indicated that we likely saw an increase in dependencies between instructions. Counter increment operations always execute sequentially, with  each instruction dependent on the prior instruction; we load an address, load a value from that address, increment it, and store it back. The total number of instructions executed also greatly increases in both cases, as measured by perf.

These stats indicate that the workloads have a lot of instrumentation on the commonly taken paths. On examining the ROI, we found that the instrumented functions had many if-else control flow statements. 
Control-flow heavy functions tend to have small basic blocks, in terms of instruction count, as basic blocks always end at control flow instructions.
\projectName may add a counter per basic block if it is unable to compute post-domination sets for the ROI. Additionally, such ROI tend to be short-lived and frequently executed, which greatly increases the impact of timing events on execution time. Furthermore, \projectName's sampling checks further complicate control flow, adding control dependencies and hurting branch predictor performance for such workloads.

This combination of factors -- adding many counters to small basic blocks for short-lived ROI is a pathological case for \projectName that can result from code with a lot of control flow.  We  note that control-flow-heavy workloads like these are unlikely to perform well on or be offloaded to accelerators. Developers using \projectName for memory bound systems are unlikely to ever encounter such cases.

On the flip side, we see lbm, which exhibits very low runtime overhead.
The ROI selected for lbm consisted of a large work loop with very little control flow.
\projectName only needed to add 4 counters and could statically determine the loop trip count,
allowing for instrumentation to be hoisted outside the loop,
and the trip count to be used to update software counters associated with it. 
The amount of instrumentation added is negligible, so the instrumented binary essentially takes as long as an uninstrumented binary. The rest of the workloads fall somewhere between these two extremes.

\begin{figure}[t]
    \centering
    \vspace{0.5 mm}
    \includegraphics[width=\linewidth]{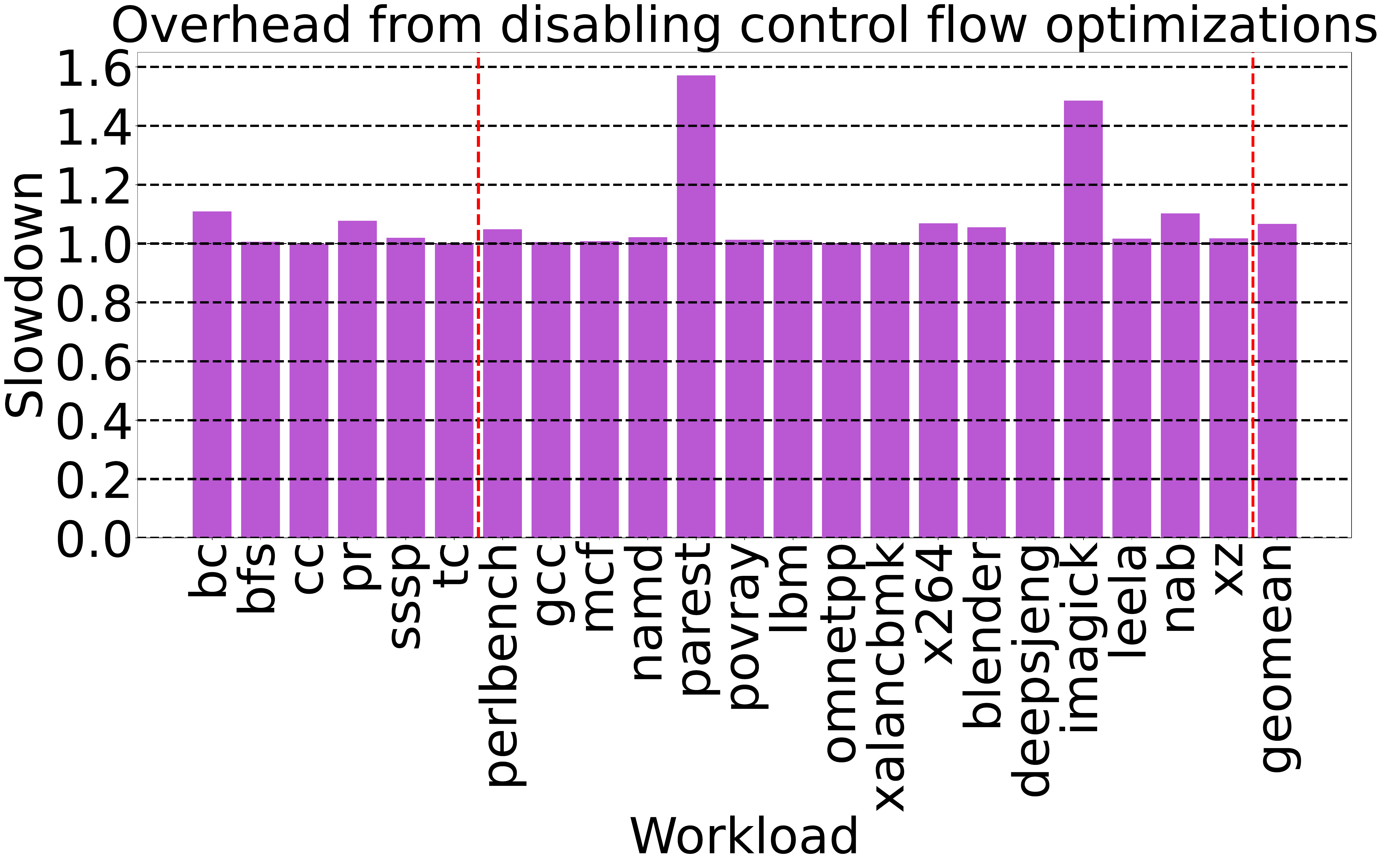}
    \caption{Runtime overhead observed when disabling loop optimizations as compared to \projectName binaries with all optimizations enabled. Vertical dotted red lines demarcate different benchmark suites. The last bar represents geomean across both benchmark suites.  
    }
    \label{fig:vii-loop-opt-overhead}
\end{figure}

We incorporate optimizations like loop hoisting and  post-domination sets to reduce the number of counters allocated and incremented by \projectName. Figure \ref{fig:vii-loop-opt-overhead} shows the performance overhead incurred by disabling these overhead when instrumenting all ROI. Most workloads see small increases in overhead, in the range of 5-10\%, but parest and imagick are notable outliers. The ROI instrumented in both cases predominantly consist of loops. The number of counters allocated and increment operations inserted in code increase by almost 30\% for both workloads.

Our sampling mechanism determines whether all counters must be incremented and whether time stamps be recorded at the start of each region of interest. As shown in Figure \ref{fig:iii-x86-overhead}, most workloads see steep drops in overhead on sampling 1\% of ROI executions, with mcf and povray dropping from taking \textgreater 3x longer (when recording information for each and every ROI execution) as an uninstrumented binary to taking only 1.2x longer. 

\begin{figure}[t]
    \centering
    \includegraphics[width=\linewidth]{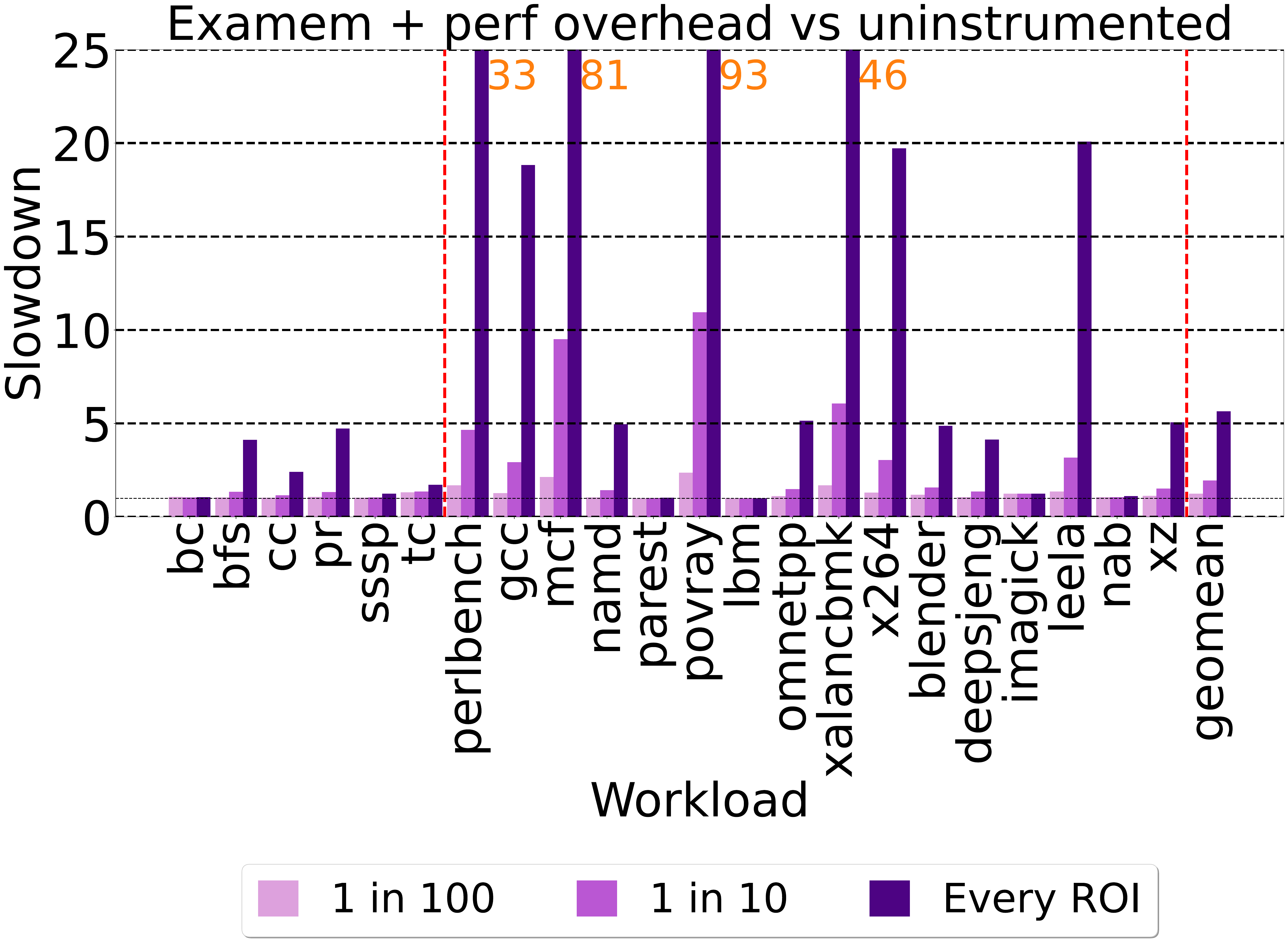}
    \caption{Runtime overhead observed with binaries instrumented using \projectName and perf compared to uninstrumented binaries. Lower is better - a value of 1 indicates no overhead. Vertical dotted red lines demarcate different benchmark suites. The last two bars represent geomean across both benchmark suites.  
    }
    \label{fig:vi-x86-perf-overhead}
\end{figure}

Additionally, we examine the overhead for both benchmark suites when instrumented with \projectName with performance counters enabled. We record three hardware performance counter events using the perf\_event library, reading the counters six times per ROI call. As can be seen in Figure \ref{fig:vi-x86-perf-overhead} recording these events adds considerable overhead. Workloads like mcf and povray, which saw a slow down over 3x with \projectName are now over 80x slower. As previously discussed, the instrumented ROI in these workloads are called very frequently and involve executing very few instructions. While \projectName's core functionality added a small amount of overhead with its counters and timing events, 6 hardware counter reads are very expensive operations that end up completely dominate the execution. In our studies, we found that on average, a read to a hardware performance counter via perf on a modern x86 machine could take upwards of 150 ns.

Given this high overhead of accessing hardware counters, they are best used with sampling. Platform-specific handlers that directly access performance counters (e.g., rdmsr/wrmsr on x86-64) could be used to improve performance further.

We examined the efficacy of sampling by studying the fraction of bytes accessed with different sampling rates as compared to when every basic block instrumented and when capturing timing information for every ROI access. Sampling is usually effective, resulting in \projectName capturing a fraction of bytes proportional to the sampling rate, and does not affect bandwidth accuracy. We saw three notable cases where byte counts were off - bc, mcf and imagick. 

In all cases, this is because of the ROI selected. For instance, our instrumentation covers the main work loop which is only executed once in bc. 
As per our current policy, an ROI always has instrumentation enable the first time it executes. Thus, for such ROI, which are executed very few times, having instrumentation enabled just that one time is enough to capture the bulk of the information, rendering the sampling policy ineffective. We conclude our study on sampling efficacy with the observation that the current sampling policy is fairly effective, but needs to be made more robust to handle cases when ROI are executed very few times.
\mnote{Can cut some of this detail for space if needed}\anote{Done}

\begin{figure}[t]
    \centering
    \includegraphics[width=\linewidth]{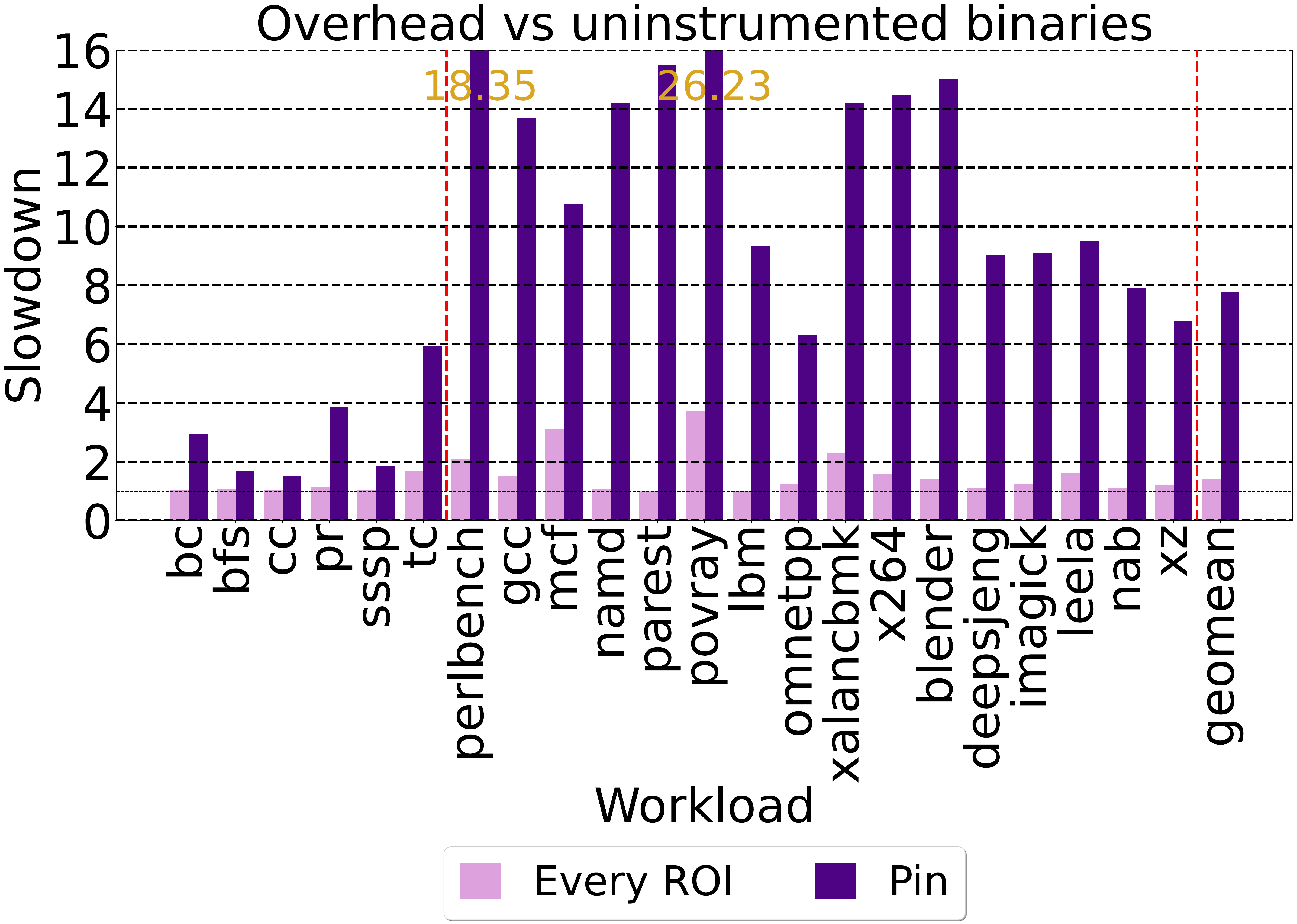}\
    \caption{Runtime overhead observed with binaries instrumented using \projectName and our pintool compared to uninstrumented binaries. Lower is better - a value of 1 indicates no overhead. Vertical dotted red lines demarcate different benchmark suites. The last two bars represent is geomean across both benchmark suites. 
    }
    \label{fig:iii-x86-overhead-pin}
\end{figure}

We also compare our performance when instrumenting every ROI execution against our pintool as seen in Figure \ref{fig:iii-x86-overhead-pin}. Geomean average runtimes for the pintool were 5x as long those taken by \projectName even in its least performant configuration.

\subsection{Portability}

\begin{figure}
    \centering
    \includegraphics[width=\linewidth]{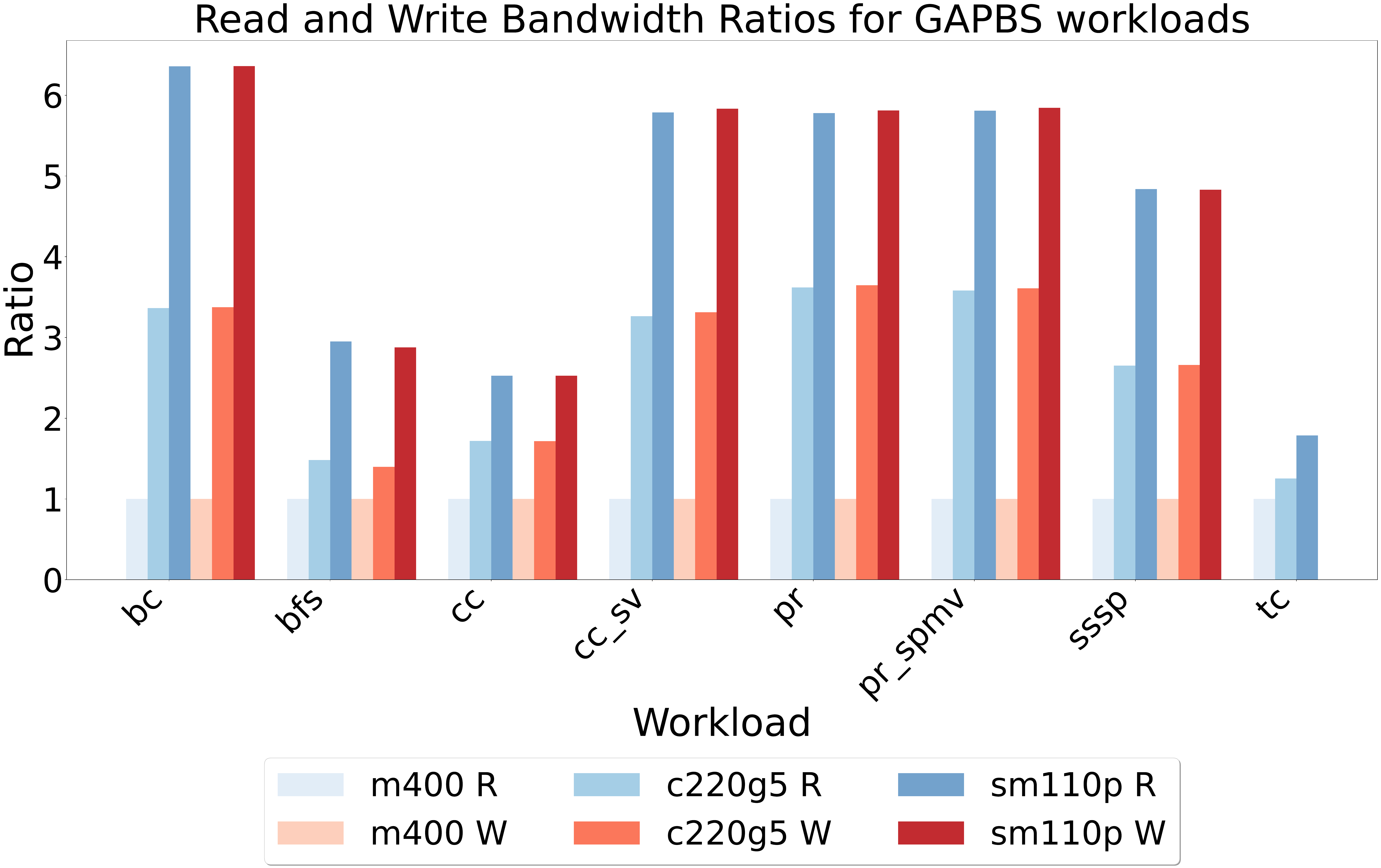}
    \caption{Ratio of heap access bandwidth for the three servers, with the m400 AArch64 server as baseline.}
    \label{fig:aarch64-gapbs}
\end{figure}


Portability is a major goal for \projectName. We evaluate the ease of bringing
\projectName to a new ISA by validating the tool on an AArch64 server. The IR pass,
runtime, and post-processing tool are all platform independent. We were able to build the IR Pass and directly use it with the package manager's version of clang, with a single change to compiler flags.\mnote{Say something about RiscV - even if that our machine is too small to run these workloads}\anote{Hayden has his board, he is looking into it. We'll add a line stating that we followed a similar approach to what we did for aarch64 and got it working on riscv (I assume it'll work because we previously got it working,) but could only run the workloads at toy sizes.}

\projectName determines the number of heap accesses and bytes accessed in IR, which is target independent. Thus, we would expect the number of accesses and bytes accessed for a workload to be similar across ISAs. We first decided to verify that this was the case. We disabled SMT and restricted OpenMP to 8 threads on the x86-64 servers (a c220g5 SkyLake server and an sm100p Ice Lake server) for parity with the AArch64 server, and confirmed that \projectName reported very similar numbers on all servers for all GAPBS workloads. 

We then compared its read and write bandwidth as reported by \projectName against those reported on the two x86-64 servers. As seen in Figure \ref{fig:aarch64-gapbs}, the latest Ice Lake server performs best as one would expect, with its larger caches and faster RAM. Triangle counting (tc) and BFS have some of the lowest LLC access and miss rates amongst the GAPBS suite and are more sensitive to cache size and performance. In case of these two workloads, we note that the AArch64 server closes the gap to the c220g5 machine, even though it uses DRAM that is a generation older, and smaller L2 caches per core, by virtue of having similar L1 cache capacity. Once again, we do not report ratios for store bandwidth for triangle counting, all machines did not report any stores.

\anote{Mentioned RISC-V and POWER8}We also tested \projectName's functionality on an IBM POWER8 server and a RISCV SBC. In both cases, setup was straightforward, and we were directly able to use \projectName's IR Pass with clang installed from the package manager and could build and run all GAPBS workloads with \projectName's instrumentation.

\section{Related Work}
\documentclass[../main.tex]{subfiles}
\graphicspath{{\subfix{../images/}}}

\begin{document}






\subsection{Performance Analysis Tools}
\anote{Commented out a small paragraph here}

Gprof~\cite{Gprof}, XRay\cite{XRay}, and a host of other tools offer low overhead profiling, but only record information at the function call level. 
Pin~\cite{Pin} allows the user to design a custom tool to collect whatever specific information they require. However, Pin only supports a few ISAs and incurs relatively high overhead.
Valgrind~\cite{Valgrind} provides features for memory debugging and profiling, but also has high overhead. 
DynInst~\cite{DynInst} can be used to dynamically instrument binaries without recompilation or re-linking. However, it requires substantial porting for new architectures and had overheads closer to Pin. \mnote{How about simplifying: dyninst requires substantial porting for new architectures and has overheads closer to PIN through excessive instrumentation?}

The previously mentioned tools are either too coarse grained in their instrumentation, suffer from high overhead, lack portability, or posses some combination of mentioned traits. LLVM-based tools maintain portability by operating at the IR level, but preexisting tools still do not meet our use-case requirements.
PISA~\cite{PISA} and its follow-up work PISA-NMC~\cite{PISA-NMC} use the LLVM infrastructure for hardware-agnostic instrumentation of workloads. PISA provides information about instruction level parallelism, memory access patterns and branch entropy. PISA-NMC extracts information about memory entropy, data reuse distance, data-level parallelism and basic-block level parallelism for near-memory computing architectures. However, such instrumentation on average resulted in an increase in execution time by two to three orders of magnitude, with the designers intending users to perform analysis only once per application and dataset. 
DINAMITE~\cite{miucin2016dinamite} is an LLVM-based memory profiling tool 
which logs memory accesses, models cache behavior, and can be used to identify potential memory bottlenecks due to poor data layout or
multi-threaded contention. However, DINAMITE's instrumentation incurs overhead similar to PIN, which interferes with our use case in
estimating achieved memory bandwidth.
LLVM Machine Code Analyzer (llvm-mca) uses processor models to simulate and calculate ROI performance and resource
pressure at the assembly level. However, llvm-mca ROI simulations do not account for program logic outside the ROI, or thread contention, that may influence actual memory performance.
\mnote{Is there an llvm-based basic-block profiler? Or any other basic-block profilers to cite?}

\subsection{Near-Memory Systems}
\projectName could be used as a performance analysis tool for code offloaded to a near-memory system. 
Existing research proposes many such systems. We envision a system with a server offloading functions or blocks of instructions to a near-memory system with general purpose cores and accelerators to take advantage of HBM. Zhang et al.~\cite{SDAM,MEG} propose systems similar to what we describe. Tian et al. ~\cite{ABNDP}  discuss the integration of general purpose cores with stacked memory. Similarly, work by Ahn et al.~\cite{PIM-Enabled-Instructions} describe a server offloading code to a PIM system with stacked memory. 

\subsection{Programming Models and Code Offload}
\hnote{May be able to remove this section if PIM/PNM is not the primary focus
and we need more space?} \anote{I think keeping at least the second paragraph is something to consider. If we want to reduce the NMP stuff, we can cut from the paragraph before this one, which is entirely devoted to that class of device.}

Processing in- and near-memory systems are currently not available for purchase to the mass market yet. Thus, we see a lot of diversity in potential system specifications and their programming models as proposed in academic research. \projectName has key functionality for instrumenting offloadable code. Boroumand et al.~\cite{PIM-workload-driven-perspective} describe four granularities for offloading code to PIM systems: applications~\cite{TETRIS}, functions~\cite{Google-Workloads-Consumer,CoNDA,SpaceA}, bulk operations, and individual instructions~\cite{PIM-Enabled-Instructions,CAIRO}.

Other methods, such as that proposed by Pattnaik et al.~\cite{PIM-GPU-Scheduling} for GPUs may also be relevant depending on the type of system and workload. Approaches for automating the process of identifying offload candidates ~\cite{TOM,Decentralized-offload} have been discussed prior. These compiler-based techniques could be used in conjunction with \projectName's compiler pass to transparently identify and instrument code that is to be offloaded without any programmer intervention.

\section{Conclusion}
\documentclass[../main.tex]{subfiles}
\graphicspath{{\subfix{../images/}}}

\begin{document}
\mnote{Update - no PIM/NMP}
Taking full advantage of the memory system requires performance analysis tools for programmers to understand application behavior. The rich set of tools available on server CPUS may not be available in virtualized environments or on the processors used in PIM or NMP systems, making it difficult to understand whether code fully uses the available memory performance. This work presents \projectName, a novel target-independent lightweight memory performance introspection framework. We make use of a basic block profiler to statically record instruction mix and insert instrumentation for regions of interest (ROI) in code, and use runtime timing and execution counts to calculate achieved bandwidth during post processing. We verify that \projectName's instrumentation is accurate and has low overhead on an x86-64 system, and demonstrate its portability by using it on 4 different ISAs and comparing the results reported for three servers. \mnote{And any other}

\section{Acknowledgments}
\documentclass[../main.tex]{subfiles}
\graphicspath{{\subfix{../images/}}}

\begin{document}
This paper was supported by the CRISP center in JUMP 1.0 and the PRISM center in JUMP 2.0, Semiconductor Research Corporation (SRC) programs sponsored by DARPA.


\bibliographystyle{IEEEtranS}
\bibliography{main}


\end{document}